\documentclass[epj]{svjour}
\usepackage{graphicx}

\begin{document}

\title{Shell Model Calculations with Modified Empirical Hamiltonian in
$^{132}$Sn region }

\author{Sukhendusekhar Sarkar\inst{1} \and M.  Saha Sarkar\inst{2} }
\mail{mss@anp.saha.ernet.in}
\offprints{M. Saha Sarkar}

\institute{
Department of Physics, Burdwan University,
Golapbag, Burdwan 713104, India.
\and
Saha Institute of Nuclear Physics,
1/AF Bidhan Nagar, Kolkata -- 700064, India.}

\date{Received: date / Revised version: date}

\abstract{
Using  recent experimental information for $^{132}$Sn region, an 
empirical Hamiltonian  is  obtained by some  modifications  of  a
Hamiltonian  (CW5082)  originally  derived  from  the  $^{208}$Pb
region. Shell model calculations with the new  Hamiltonian  shows
remarkable improvement in the predictive power when compared with
the   available   experimental  results. It overcomes many limitations
of  the  CW5082  Hamiltonian  in  this region,  specially  for 
N$\geq$84 isotones. The calculated level
spectra and B(E2) values with the new Hamiltonian, also  compare
well  with  the available results calculated with the CD-Bonn and
SKX Hamiltonians, reflecting consistency  in  the  wave  function
structure   at   least  in  the  low-lying  regions.  Interesting
behaviour of effective charges is  revealed in this region. It is
shown that a drastic reduction  of  proton  effective  charge  is
necessary  for  obtaining  B(E2)  values  for  the N=84 isotones.
Neutron effective charge is found to be  in  the  range  (0.62  -
0.72)e. We predict the spectroscopic properties of $^{135,136}$Sn
not yet known experimentally. Further improvement of the modified 
Hamiltonian is also initiated.}

\PACS{
      {21.60.Cs}{Shell model}  \and
      {21.10.-k}{Properties of nuclei, nuclear energy levels} \and
      {23.20.-g}{Electromagnetic transitions}
      {27.60.+j}{ 90 $\leq$ A $\leq$ 149}
     }

\titlerunning{Shell Model Calculations in the$^{132}$Sn Region}
%\titlerunning{Interesting Features of Effective Charges Revealed}

\maketitle

\section{Introduction}
\label{intro}
Few  -  valence  -  particle  neutron  -  rich  nuclei  above the
strongest doubly closed $^{132}$Sn \cite{Blo:1}  are  interesting
for  many  reasons. They provide opportunity to extract empirical
N-N interaction, as well  as  to  test  theoretical  shell  model
description  of  nuclear  structure  in  this  region.  Structure
properties of some of  these  nuclei  are  important  inputs  for
astrophysical r - process model calculations.

In  a  previous attempt \cite{Sah:1}, we have studied some N=82 -
85 isotones in this region with  two  (1  +  2)  -  body  nuclear
Hamiltonians,  namely,  KH5082  and CW5082, using the shell model
code OXBASH \cite{Oxb:1}. The KH5082 and CW5082 are described  in
detail  in  the  work  of  Chou  and  Warburton \cite{Cho:1}. One
important observation of our study \cite{Sah:1} was that both the
interactions, especially the CW5082  worked  reasonably  well  in
predicting  binding  energy,  level  spectra  and electromagnetic
properties of nuclear states for some N = 82 - 83 isotones in the
$^{132}$Sn region. But for N$\geq$ 84 isotones, the agreement  of
the  calculated  values with those from experiments were poor. We
have pointed out that, particularly, the neutron  ($\nu$)-  $\nu$
two-body  matrix  elements (tbmes) of these interactions might be
inappropriate and this has also been supported by some other very
recent    experimental     and     theoretical     investigations
\cite{Rad:1,Cor:1}.  Thus KH5082 and thereby CW5082 Hamiltonians,
obtained by $A^{-1/3}$ scaling from Kuo - Herling Hamiltonian for
the stable $^{208}$Pb region after some  realistic  modification,
though  works  for the $^{132}$Sn region within limitations, must
be changed on account of the relatively large neutron -  richness
in  the  $^{132}$Sn  region. Sn, Sb isotopes above the $^{132}$Sn
core  are  already  about  10  -  11  neutrons  away  from  their
corresponding last stable isotopes.

We  have  initiated  \cite{Dae:1}  to  modify the existing
nuclear Hamiltonian CW5082 in the light of the  recent  data  for
this  region.  We  have  obtained remarkable improvement with a
modified interaction by changing a few neutron - neutron and neutron
- proton matrix elements and report in  this  article  some  of  the
interesting results. Further improvement in the modified Hamiltonian is
indicated by changing also a few proton-proton matrix elements.

\section{Formalism: Model space and modified Hamiltonian}
\label{sec:1}

It  has  been  mentioned  above  that the CW5082 predicts binding
energies,   level    spectra    and    other    properties    for
50$\leq$Z$\leq$55  and  $N< 84$ nuclei reasonably well. So in the
present attempt, we have  modified  the  CW5082  interaction.  We
assume  $^{132}$Sn  as the inert core. The valence space consists
of  five  proton   orbitals,   $\pi$   [$1g_{7/2}$,   $2d_{5/2}$,
$2d_{3/2}$,  $3s_{1/2}$,  $1h_{11/2}$]  and  six neutron orbitals
$\nu$   [$1h_{9/2}$,    $2f_{7/2}$,    $2f_{5/2}$,    $3p_{3/2}$,
$3p_{1/2}$,   $1i_{13/2}$],  with  energies  in  MeV,  $\pi$  [0.
(-9.6629) \cite{Fog:99},  0.9624,  2.4396,  2.6972,  2.7915]  and
$\nu$  [1.5609,  0.0  (-2.4553)  \cite{Fog:99},  2.0046,  0.8537,
1.6557, 2.6950], respectively. Here  we  have  replaced  all  the
single  particle  energies  (spes)  of  the  proton  and  neutron
orbitals in CW5082 by experimentally determined ones \cite{Bnl:1}
(except $\pi 3s_{1/2}$ and $\nu 1i_{13/2}$ spes). $\pi  3s_{1/2}$
spe  is obtained from the local systematics \cite{Cho:1} and $\nu
1i_{13/2}$ is taken from Urban et al. \cite{Urb:1}.

Five proton- neutron tbmes of CW5082 interaction were obtained by
Chou and Warburton by adjusting them to reproduce the energies of
$I^\pi  = 0^-$ and $1^-$ levels of $^{134}$Sb. It is important to
note that a recent precise measurement of the binding  energy  of
this  $0^-$  state  by  Fogelberg  et al. \cite{Sah:1,Fog:99} has
changed the previous value \cite{Wap:97} by a significant  amount
of  about  200  keV.  The recently measured \cite{Fog:99} binding
energy of $^{132}$Sn is also different  from  the  earlier  value
\cite{Wap:97}.  So  the $\nu-\pi$ tbmes also need modification to
incorporate these important changes.

We   change  the  neutron-neutron  and
proton-neutron tbmes keeping the proton-proton tbmes the same  as
those   in   CW5082.   The  six  $\nu-\nu$  diagonal  tbmes  with
$I^\pi=0^+$ were already noted to be too attractive in  the  work
of  Chou  and  Warburton  \cite{Cho:1}. We multiply all these six
tbmes by a factor of 0.48. This factor is obtained by reproducing
the binding  energy  of  $^{134}$Sn  (-6.365).  All  the  binding
energies  (in  MeV) are with respect to $^{132}$Sn \cite{Fog:99}.
Three excited states in  $^{134}$Sn  \cite{Kor:1},  predominantly
from the $(\nu 2f_{7/2})^2$ multiplet, at energies 725.6, 1073.4,
and 1247.4 keV are used to modify the $\langle{ {\nu 2f_{7/2}}^2|
V| \nu 2f_{7/2}}^2\rangle^{2^+,4^+,6^+}$ tbmes. The $\langle {\nu
1h_{9/2}  2f_{7/2}}|  V| \nu 1h_{9/2} 2f_{7/2}\rangle^{8^+}$ tbme
has been changed to reproduce the energy of $8^+$ level at 2508.9
keV.

Similarly,  using  binding  energy  (-12.952)  and  $1^-$, $2^-$,
$3^-$, $4^-$, $7^-$, $8^-$, $10^+$,  $9^+$,  $10^-$,  $11^-$  and
$12^-$  excited  levels  at  energies  13.0, 330.7, 383.5, 554.8,
283.0 \cite{Sah:1}, 1073, 2434, 2126, 4094,  4425  and  4517  keV
respectively,  of  $^{134}$Sb  \cite{Sb:1},  we  have modified 12
dominant proton-neutron tbmes. Thus  we  have  changed  only  ten
$\nu-\nu$  tbmes  and  twelve  $\nu-\pi$  tbmes to obtain SMN5082
Hamiltonian from CW5082.

From   the  results  of  shell  model  calculations  with  CW5082
interaction it  was  found  that  its  proton-proton  tbmes  were
reasonable.  So  in  obtaining  SMN5082 the $\pi- \pi$ tbmes have
been not  changed.  Recently,  the  two  valence  proton  nucleus
$^{134}Te$   \cite{Te:1}   has  been  studied  more  extensively,
compared to $^{134}Sn$ and $^{135}Sb$ nuclei. It is expected that
inclusion of new experimental data  of  $^{134}Te$  will  further
improve  the interaction (SMN5082). We have initiated this effort
by using a few $^{134}$Te data. We have used the  binding  energy
of  $^{134}$Te (-20.56) and its three excited levels $2^+$, $4^+$
and $6^+$ at  energies  1279,  1576  and  1692  keV  \cite{Te:1},
respectively,   predominantly   from   the   $(\pi   1g_{7/2})^2$
multiplet, to modify only four  important  $\pi-  \pi$  tbmes  of
SMN5082 and the resulting interaction is named as SMPN5082.

We  shall  refer to, for brevity, SMN5082 and SMPN5082 as SMN and
SMPN, respectively.  Thus,  of  the  2101  tbmes  of  the  CW5082
Hamiltonian, we have changed only 22 tbmes for SMN and additional
four proton-proton tbmes for SMPN.

\subsection{Results and Discussions}
\label{sec:2}

With these new interactions, we have calculated binding energies,
level  spectra and B(E2) values for some N=84-87 Sn, Sb, Te and I
isotopes. Two different calculations have  been  carried
out  with  these  two slightly differing interactions. It is seen
that even with this minor change, the result for the binding energies
with   the   SMPN interaction  is  consistently  better  than  that
of  SMN. So we compare  the  binding   energies   calculated   with   both   the
interactions.  It  should be noted that the results for Sn and Sb
isotopes should be  the  same  for  both  the  interactions.  For
$^{136}$Te  and $^{137}$Te, $^{137}$I the results are expected to
be slightly different with SMN and SMPN. The level  spectra  with
the  SMN  interaction  for  $^{135}$Sb  and  $^{137}$I  have been
compared in (fig (a),(c)) with the experimental as well  as  that
of  the parent CW5082, to show the improvement. In fig(b) and (d)
we give for comparison the experimental level spectra  and  those
calculated with SMN and SMPN interactions for $^{136-137}$Te.

It  is  emphasised that the new SMN and SMPN interactions produce
consistently good results also for the N=82-83 isotones of  these
and  Xe, Cs nuclei, which we have studied earlier with CW5082 and
KH5082 \cite{Sah:1}.

The calculated binding energies (table 1) can be largely affected
if the spes contain errors. The $\pi 3s_{1/2}$ and $\nu i_{13/2}$
spes  might  have  uncertainties  because  of  the  ways they are
determined. These have small effect for the ground states of  the
nuclei  considered.  The  binding  energies  (table  1)  are well
reproduced with the two new Hamiltonians for N = 82, 83 isotones.
For N$\geq$85, Sn and Sb nuclei, the binding energies  quoted  as
experimental are derived from the local systematics and therefore
have large errors (table 1). Therefore, it is difficult to draw a
definite  conclusion  regarding  the  agreement of the calculated
binding energies with experimental values for these  nuclei.  But
for  N=84 isotones and $^{137}$Te$_{85}$, slight over-binding can
be  noted  for  the  calculated  values.  So   we   assume   this
over-binding  to  persist  for $N\ge84$, Sn and Sb isotopes also.
With the CW5082, the calculated binding energies  are  relatively
less over-bound compared to the SMN and SMPN for N$\geq$ 84. This
is  because  the  neutron  spes of CW5082 were made less bound by
adding 100 keV to each neutron  spe.  Our  observation  regarding
this  over-binding  for  $N\geq$84  is that it has an approximate
systematic property. Over-binding, with SMPN, is  $\Delta  \times
(N-82)$, $\Delta = $ 0.1 and 0.175 MeV for I and Te respectively.
For   $^{135-137}$Sb   this   is   0.4(N-83)   MeV  and  that  in
$^{135-137}$Sn, is 0.6(N-84) MeV except for even-even $^{136}$Sn.
The over-binding decreases with increasing valence proton number.
This systematics as well as the  detailed  consideration  of  the
interacting  $\pi-\pi$,  $\pi-\nu$  and  $\nu-\nu$  pairs  in the
valence space  of  the  $N\geq  84$  isotones  suggest  that  the
over-binding    has    a    connection    with   the   increasing
neutron-richness  or  $N/Z$  ratio.  The  over-binding  increases
particularly  with the increasing $\nu-\nu$ pairs. This indicates
the necessity of modification of  more  $\nu-\nu$  and  $\nu-\pi$
tbmes beyond the dominant ones, which requires further data.

In  Figs.  1(a-d),  we  compare excitation spectra from our shell
model calculations with CW5082, SMN and  SMPN  Hamiltonians  with
very  recent  experimental  spectra.  The  agreement  for all the
nuclei is excellent, except for the  282  keV  $5/2^+$  level  in
$^{135}$Sb,   showing   improvement   achieved  through  the  new
Hamiltonians.

We  compare our results for $^{135}$Sn, for which no experimental
spectra is available yet, (except estimates from systematics  for
lowest  few  levels  by  Urban  et  al.  \cite{Urb:3}), with that
predicted in Ref. \cite{Cor:1}. The energies with respect to  the
$7/2^-$  ground state of the $5/2^-$, $3/2^-$, $11/2^-$, $9/2^-$,
$15/2^-$, $3/2^-_2$, $9/2^-_2$ and $7/2^-_2$ levels are  at  233,
353,  657,  701, 993, 1020, 1434 and 1535 keV in our calculation,
whereas these levels are at 226, 356, 611, 706,  911,  643,  1093
and  1192 keV, respectively, in Ref. \cite{Cor:1}. The energies of
the first five excited states of our calculation compare  closely
to the corresponding results of Ref. \cite{Cor:1} as well as with
the estimates by Urban et al. \cite{Urb:3}. But the result starts
deviating by more than 200 keV for higher excited levels with the
calculation of Ref. \cite{Cor:1}.

Encouraged  by  the  good  agreement  of  our  results  with  the
available experimental data for all the nuclei, we  predict  also
the spectra of $^{136}$Sn, an important nucleus for the r-process
nucleosynthesis.  The  excitation  energies  with  respect to the
$0^+$ ground state of the $2^+$, $4^+$, $4^+_2$, $6^+$,  $2^+_2$,
$5^+$,  $8^+$  and  $2^+_3$ levels are 578, 886, 994, 1086, 1106,
1272, 1682 and 1696  keV,  respectively.  It  is  interesting  to
compare  our prediction for the energy of the first excited $2^+$
state at 578 MeV with the estimate for it at about 600  keV  from
systematics by Urban et al \cite{Urb:3}.

The  excitation  spectra  of $^{135}$Sb (Fig. 1a) studied through
prompt  gamma  ray  emission  up  to  $23/2^+$  \cite{Pbh:1}   is
reproduced  excellently  in our shell model calculation using the
new  SMN  Hamiltonian.  But  the  $5/2^+$  level   at   282   keV
\cite{She:1}  populated  via  beta  decay  of  $^{135}$Sn  is not
reproduced in our  calculation  ($E_{5/2^+}$  =  690  keV).  This
behaviour  is  also noticed in other shell model calculations and
have been discussed elaborately by Shergur  et  al.  \cite{She:1}
and references therein.

In  $^{136}$Te  a  long  lived isomeric $12^+$ level was expected
\cite{Hof:1} from the analogy  with  $^{212}$Po,  as  well  as  from
theoretical  calculations  \cite{Kor:2}.  But  it  was  not  seen
experimentally.  Its  non-existence  was  confirmed  \cite{Kor:2}
experimentally   in  an  indirect  way  from  the  study  of  the
neighbouring  $^{137}$I  nucleus.  Our   results   in   Fig.   1b
excellently  reproduces the experimental data confirming that the
observed $12^+$ level at 3187 keV is indeed yrast and non-isomer.
Thus  the  missing  $12^+$  isomer  issue  \cite{Kor:2}  is  also
resolved  theoretically.  The  $2^+_2$  level  in this nucleus is
estimated at 1568 keV from systematics \cite{Hof:1}, whereas  our
result is 1591 keV (SMPN) and 1603 keV (SMN). The position of the
lowest  $3^-$  level  in  $^{136}$Te,  which  is  of considerable
interest \cite{Hof:1} is also shown in Fig. 1b.  to  be  at  3284
keV.  This level, obviously involves many more unchanged tbmes of
the original CW5082 and thus may have  some  uncertainty  in  the
calculated energy.

The spectra of $^{137}$I \cite{Kor:2} and $^{137}$Te \cite{Urb:2}
(Figs. 1c,d) show a kind of regularity indicating collectivity in
their  excitations.  Clear  indication  of signature splitting is
seen  in  the  spectra  of  both  $^{137}$I,   Te   nuclei.   Our
calculations  reproduce  this  behaviour perfectly for $^{137}$I.
But for $^{137}$Te, all the levels with  signature $-1/2$ (except
for $31/2^-$) are reproduced within $\approx 50$ keV, whereas the
levels with +1/2 signature deviate by $\approx 150-200$ keV.

The  structures  of  the  wave  -  functions  of $^{137}$I energy
levels  show  large  configuration   mixing   leading   to   such
collectivity.  For  example,  with  SMN,  for  the  $5/2^+$ first
excited  state,  13  configurations  contribute   75\%   to   the
wave-function  and  the  rest is contributed by at least 25 other
configurations.   The   most   dominant   $(\pi   1g_{7/2})^3(\nu
2f_{7/2})^2$   configuration   contributes  35.5\%  to  the  wave
function. One can compare this with the structure  of  the  first
excited  state  ${5/2}^+$  in  $^{135}$I$_{82}$ for which neutron
shell is closed. Only 5 configurations contribute  98\%,  out  of
which 80\% is contributed by the $(\pi 1g_{7/2})^2(\pi 2d_{5/2})$
configuration.  This  state is therefore, predominantly of single
particle nature.

In  $^{137}$Te  there  is  an estimate of the energy of the first
${9/2}^-$\cite{Urb:3} level from the experimental systematics  at
around 700 keV above the ${7/2}^-$ ground state. This is close to
our results in Fig. 1d (802 keV). Similarly, estimated $({5/2}^-,
{3/2}^-)$  level at around 100 keV of \cite{Urb:3} is most likely
a ${5/2}^-$ as shown by our calculation in the same Fig. 1d at 97
keV. Figs. 1c and d  help  resolving  ambiguous  $I^\pi$  of  the
levels in $^{137}$I and $^{137}$Te.

From  the  very good agreement of the calculated spectra with the
experimental ones for all the nuclei compared here, one can  hope
that  the  new  Hamiltonians,  SMN  and  SMPN,  can be useful for
describing structure properties of at least low-lying  yrast  and
near   yrast  states  for  all  these  few-valence-particle  very
neutron-rich nuclei in the $^{132}Sn$ region.

To test the wave - functions corresponding to SMN Hamiltonian and
to  derive  effective charges for this region, we have calculated
the B(E2) values for the transitions $2^+ \rightarrow 0^+$,  $4^+
\rightarrow 2^+$, $6^+ \rightarrow 4^+$ and $8^+ \rightarrow 6^+$
in $^{134}$Sn and obtained the values 73.2, 73.8, 36.5 and 4.8 in
e$^2$fm$^4$,   respectively,   with   neutron   effective  charge
$e_n^{eff}$ = 0.72e. (B(E2) values are  expressed  throughout  in
e$^2$fm$^4$).  This  value  of  effective  charge  $e_n^{eff}$  =
${{0.72}^{+0.06}}_{-0.08}$e, has been fixed  by  reproducing  the
experimental  B(E2)  value  (36$\pm7$)  of  $6^+ \rightarrow 4^+$
transition in $^{134}$Sn  \cite{Sn:1}.  The  corresponding  B(E2)
values  in  Ref.  \cite{Cor:1} with $e_n^{eff}$ = 0.70e are 70.1,
69.6, 35.8 and 4.9, respectively, showing  very  good  agreement.
Henceforth effective charges will be expressed in unit of $`e`$.

In  table  2,  we  have compared the calculated and the available
experimental  B(E2)  values  \cite{Rad:1,Sn:1}  for   $^{134}Sn$,
$^{134,136}Te$  and  $^{135}Sb$  nuclei  with  different  sets of
effective charges. For $^{135}Sb$, only approximate value for the
half-life is found in the literature \cite{Bnl:1,Pbh:1}. A B(E2)  value
$\simeq 45$ has been extracted from the value of the half-life of
the  $19/2^+$  level  using expression given in Ref. \cite{Law:1} and
theoretical  value  of  the  internal   conversion   co-efficient
\cite{Bnl:2}.

It   is   found  that  to  reproduce  the  B(E2)  value  for  the
$6^+\rightarrow4^+$ transition in $^{134}Te$ a  proton  effective
charge   as   low   as  $1.34\pm0.01$  is  needed.  Whereas,  for
$0^+\rightarrow2^+$ transition the corresponding proton effective
charge is about $1.54\pm 0.10$. It is important to note that  the
proton  effective  charge of $6^+\rightarrow4^+$ transition gives
B(E2)  value  for   the   $0^+\rightarrow2^+$   less   than   the
experimental  lower  limit  for  it.  The  B(E2)  value  for  the
$6^+\rightarrow4^+$ transition appears to be  more  precise  than
the  $0^+\rightarrow2^+$  transition.  With  $e_p^{eff}$=1.47, (a
value  in  between,  and  used   in   literature   \cite{Wil:1}),
$\hbar\omega  =  45A^{-1/3} - 25A^{-2/3}$ as in Ref. \cite{Sah:1},
the calculated B(E2) values for the  $6^+  \rightarrow  4^+$  and
$0^+ \rightarrow 2^+$ transitions in $^{134}$Te (table 2) compare
well with the experimental values.

However, the B(E2) values with $e_n^{eff}$ = 0.72 and $e_p^{eff}$
=  1.47, for the $19/2^+ \rightarrow 15/2^+$ and $0^+ \rightarrow
2^+$  transitions  in $^{135}Sb$ and $^{136}Te$, are 84 and 2165,
respectively.  These   are   about   double   the   corresponding
experimental  values  45  and  1030.  This kind of result is also
obtained with other Hamiltonians (CD-Bonn and SKX) and have  been
discussed  by  Radford  et  al.  \cite{Rad:1}, and Shergur et al.
\cite{She:1}. Recently, problems close  to  this  has  also  been
studied  by  Terasaki  et  al. \cite{Ter:1} in a schematic model.
They attributed the anomalous behaviour  of  the  $0^+\rightarrow
2^+$  E2 transition in the $^{136}$Te to the reduction in neutron
pairing  above  the  N=82  magic  gap.  Thus  effective   charges
extracted   from   the   pure  proton  ($^{134}$Te)  and  neutron
($^{134}$Sn) systems can not reproduce the B(E2) values  for  the
N=84, Sb and Te nuclei.

Now,  let  us  consider  $^{134}$Sn,  $^{134}Te$  and $^{136}$Te,
keeping aside the N=84, $^{135}$Sb nucleus. For $^{136}Te$, if we
use  the  proton   effective   charge   1.47   from   $^{134}Te$,
${e_n}^{eff}$ needed is $\simeq$ 0. This value is anomalously low
compared  to  ${e_n}^{eff}$ (0.72) for $^{134}$Sn. Similarly, one
can use ${e_n}^{eff}$ from $^{134}$Sn and vary  ${e_p}^{eff}$  to
fit  B(E2)  value  of  $^{136}$Te. In this case the ${e_p}^{eff}$
$\simeq$ 0.80, which is anomalously low compared to that  derived
for $^{134}$Te.

But,  if we include $^{135}Sb$ also and consider the B(E2) values
of three N =  84  isotones  of  Sn,  Sb  and  Te,  the  following
observations  can  be  made. We find that ${e_p}^{eff}$ = 1.0 and
${e_n}^{eff}$ = ${0.51 \pm 0.11}$  reproduces  the  B(E2)  value,
$1030  \pm  150$  for  the  $0^+  \rightarrow  2^+$ transition in
$^{136}Te$. The same  set  of  effective  charges  give  B(E2)  =
${42_{-13}}^{+14}$    for    the    ${19/2}^+\rightarrow{15/2}^+$
transition in $^{135}Sb$, and this is close to  the  experimental
value.  Similarly,  with  ${e_p}^{eff}$=1.0  and  ${e_n}^{eff}$ =
0.54, which reproduces very closely the measured B(E2) value  45,
for  the  ${19/2}^+\rightarrow{15/2}^+$ transition in $^{135}Sb$,
one gets a value 1070 for the $0^+\rightarrow 2^+$ transition  in
$^{136}Te$   and   is   within   the   experimental   range.  But
${{e_n}^{eff}}$ from these sets of effective charges  give  about
half  of  experimental  B(E2)  value for the $6^+\rightarrow 4^+$
transition in $^{134}Sn$, (since ${e_n}^{eff}$ =  0.72).  So  for
these  N=84  isotones, only proton effective charge $\simeq$ 1.0,
with the neutron effective charge in the  range  (0.62-0.72)  can
bring  all  the calculated B(E2) values close to the experimental
limits. It may be found from table 2 (column  7)  that  the  most
reasonable  choice for a single set of effective charges for N=84
isotones is ${e_p}^{eff} = 1.0$ and ${e_n}^{eff}  =  0.62$.  With
SMPN Hamiltonian similar results are obtained.

 So  this  drastic  reduction of proton effective charge for N=84
isotones compared to the N=82 isotones is quite  interesting  and
needs    further    detailed    experimental    and   theoretical
investigations.

In  $^{136}Sn$, the $\Delta$v (seniority) $= 0$, $6^+ \rightarrow
4^+$ transition between the  pure  $(\nu  2f_{7/2})^4$  multiplet
levels  is  expected  to  be severely inhibited, as in $^{136}Xe$
\cite{Sah:1,Law:1,Dal:1}.  With  SMN,  the   $(\nu   2f_{7/2})^4$
configuration  is although dominant in the $6^+$ and $4^+$ levels
in $^{136}$Sn (about 80\% and 75.6\%, respectively), yet  due  to
configuration  mixing  the B(E2) value for this transition is not
so  severely  inhibited.  It  is  predicted  to  be   12.9   with
$e^{eff}_n$=0.62.  Measurement  for this B(E2) value will be very
helpful to  conclude  whether  a  further  reduction  of  neutron
effective charge is needed in this N=86 Sn isotope.

\section{Conclusion}
\label{sec:3}

In  summary,  our  calculations  with  the SMN Hamiltonian
obtained by modifying CW5082 show several interesting results for
this very neutron - rich region. We  have  found  good  agreement
with  experimental  data  for  the  excitation spectra of all the
nuclei  except  for  the  first  $5/2^+$  level  at  282  keV  in
$^{135}$Sb. Our results clearly help in assigning spin and parity
of  some  levels  in  these  nuclei.  Predictions  for  levels in
$^{135}Sn$ and $^{136}$Sn may motivate and provide a guidance for
future experiments. Results for $^{136}Te$ resolve  theoretically
the  missing  $12^+$  isomer  issue  and  predict the position of
${3_1}^-$ level in it. The calculations for the A =  137,  I  and
Te,  reveal  signature  of  collectivity. Calculated B(E2) values
with usual e$^{eff}_p$ around 1.47 for $^{134}Te$ isotope compare
well with the  very  recent  experimental  data.  But  a  drastic
reduction  in  proton  effective  charge is required to reproduce
B(E2) values for N=84 isotones. Neutron effective charge seems to
lie in the range (0.62-0.72). Further modification of SMN is initiated 
in SMPN by modifying a few $\pi-\pi$ tbmes. We are looking  forward  to  change
more $\pi-\pi$ tbmes in SMPN using some already existing data and
hope  to  modify  more  tbmes and spes using further experimental
information which will be available in near future.

\section{Acknowledgment}
\label{sec:4}

 The   authors  thank  Prof.  Sudeb  Bhattacharya  and  Prof.  B.
Dasmahapatra for encouragement.

\begin{center}
{\bf  FIGURE CAPTIONS}
\end{center}

\begin{itemize}

\item{Fig.   1}   Comparison   of   calculated  and  experimental
excitation energies for N=84,85  isotones,  (a)  $^{135}Sb$,  (b)
$^{136}Te$, (c) $^{137}I$, (d) $^{137}Te$.

\end{itemize}

\begin{table}

\caption {Comparison of calculated and experimental \cite{Fog:99}
binding energies in MeV with respect to $^{132}Sn$.}

\begin{center}
 \smallskip
 \begin{tabular}{ccccc}\\
\noalign{\hrule}
Isotope &Expt.$^a$     &\multispan{3}      \hfil      Theoretical\hfil\cr
&\cite{Fog:99}   &\multispan{3}\hrulefill   \cr  &&  SMN  &  SMPN
&CW5082\cr
\noalign{\hrule}

$^{134}Te_{82}$&20.560(26)&20.643&20.560$^b$&20.512                 \cr
$^{135}I_{82}$&29.083(25)&29.303&29.055&29.102                    \cr
$^{136}Xe_{82}$&39.003(25)&39.368&38.977&39.103                   \cr
$^{137}Cs_{82}$&46.419(24)&46.911&46.247&46.582\cr
$^{134}Sb_{83}$&12.952(52)&12.952$^b$&12.952&12.768                 \cr
$^{135}Te_{83}$&23.902(93)&23.990&23.913&23.624  \cr
$^{136}I_{83}$ &32.861(55)&32.983&32.753&32.505          \cr
$^{137}Xe_{83}$ &43.029(25)&43.482&43.124&42.863    \cr
$^{134}Sn_{84}$& 6.365(104)&6.363$^b$&6.363&6.705             \cr
$^{135}Sb_{84}$&16.565(113)&16.989&16.989&17.017                   \cr
$^{136}Te_{84}$&28.564(55)&28.975&28.907&28.860   \cr
$^{137}I_{84}$ &37.934(37)&38.336&38.131&38.011               \cr
$^{135}Sn_{85}$&8.437(401)$^c$&9.053&9.053&8.926\cr
$^{136}Sb_{85}$&19.516(301)$^c$&20.306&20.306&19.759\cr
$^{137}Te_{85}$&31.775(122)&32.410&32.345&31.762\cr
$^{136}Sn_{86}$&12.208(501)$^c$&13.041&13.041&13.162\cr
$^{137}Sb_{86}$&23.257(401)$^c$&24.477&24.477&24.144\cr
$^{137}Sn_{87}$&14.280(600)$^c$&16.046&16.046&15.179\cr
\noalign{\hrule}
 \end{tabular}
 \end{center}
 $^a$  Errors  are  within  parentheses.  $^b$  Fitted.  $^c$  From
systematics \cite{Wap:97}

\onecolumn
\caption{Comparison  of  calculated  (with  SMN) and experimental
(error not available for Sb) B(E2) values for N=84 isotones using
four sets of (proton, neutron) effective charges in  the  present
calculation.  Ranges  of  values for neutron and proton effective
charges are discussed in the text. Note that the calculated B(E2)
values  for  $0^+\rightarrow2^+$  transition  in  $^{134}Sn$   is
included for which no measurement has yet been reported.}

\begin{center}
\smallskip
\begin{tabular}{ccccccccc}\\
\noalign{\hrule}
         &  &\multispan{7}  \hfil B(E2) in ($e^2 fm^4$) \hfil \cr
Isotope & $I^\pi_i \rightarrow I^\pi_f$  &\multispan{7}\hrulefill
\cr
         && Expt.  &\multispan{6} \hfil Theoretical\hfil\cr
         &&\cite{Rad:1,Bnl:1,Pbh:1,Sn:1,Omt:95}
         &\multispan{6}\hrulefill \cr
         && & \multispan{4}\hfil SMN \hfil&CW5082\cite{Sah:1}& CD
Bonn\cite{Cor:1}\cr
         && & \multispan{4}               &                  & SKX
\cite{She:1}\cr
         &&      &$e_p^{eff}$1.47  &0.80&1.47 &1.0 &1.47
	 &1.55\cite{Cor:1}, 1.5\cite{She:1}\cr
         &&      &$e_n^{eff}$ 0.72 &0.72&0.0 &0.62&1.00&0.70\cr
\noalign{\hrule}
\cr
$^{134}_{52}Te_{82}$&  6$^+ \rightarrow 4^+$&$83.5\pm1.2$
&100.2&29.7&100.2&-&97.1 &-
\cr
                    &  0$^+ \rightarrow 2^+$&$960\pm120$
		    &869.8&257.6&869.8&-
&858.4&880\cite{Cor:1}\cr
                    &                       &             &     & &&
&     &810\cite{She:1}\cr
\cr
$^{136}_{52}Te_{84}$&  0$^+ \rightarrow
2^+$&$1030\pm150$&2165&1043&975&1185&2989&2500\cite{Cor:1}\cr
                     &
    &            &    &    &    &    &    &2300\cite{She:1}\cr
\cr
$^{134}_{50}Sn_{84}$&  6$^+ \rightarrow
4^+$&$36\pm7$&37&37&0&27&97&- \cr
    &  0$^+ \rightarrow 2^+$&    -   & 366&366&0.0&271&-&-
\cr
\cr
$^{135}_{51}Sb_{84}$& 19/2$^+ \rightarrow
15/2^+$&$\simeq 45$ & 84 & 67 & 5 & 56 & 89 &-\cr
\cr
\noalign{\hrule}
\end{tabular}
\end{center}
\end{table}

\begin{figure*}
\resizebox{0.95\textwidth}{!}{
  \includegraphics{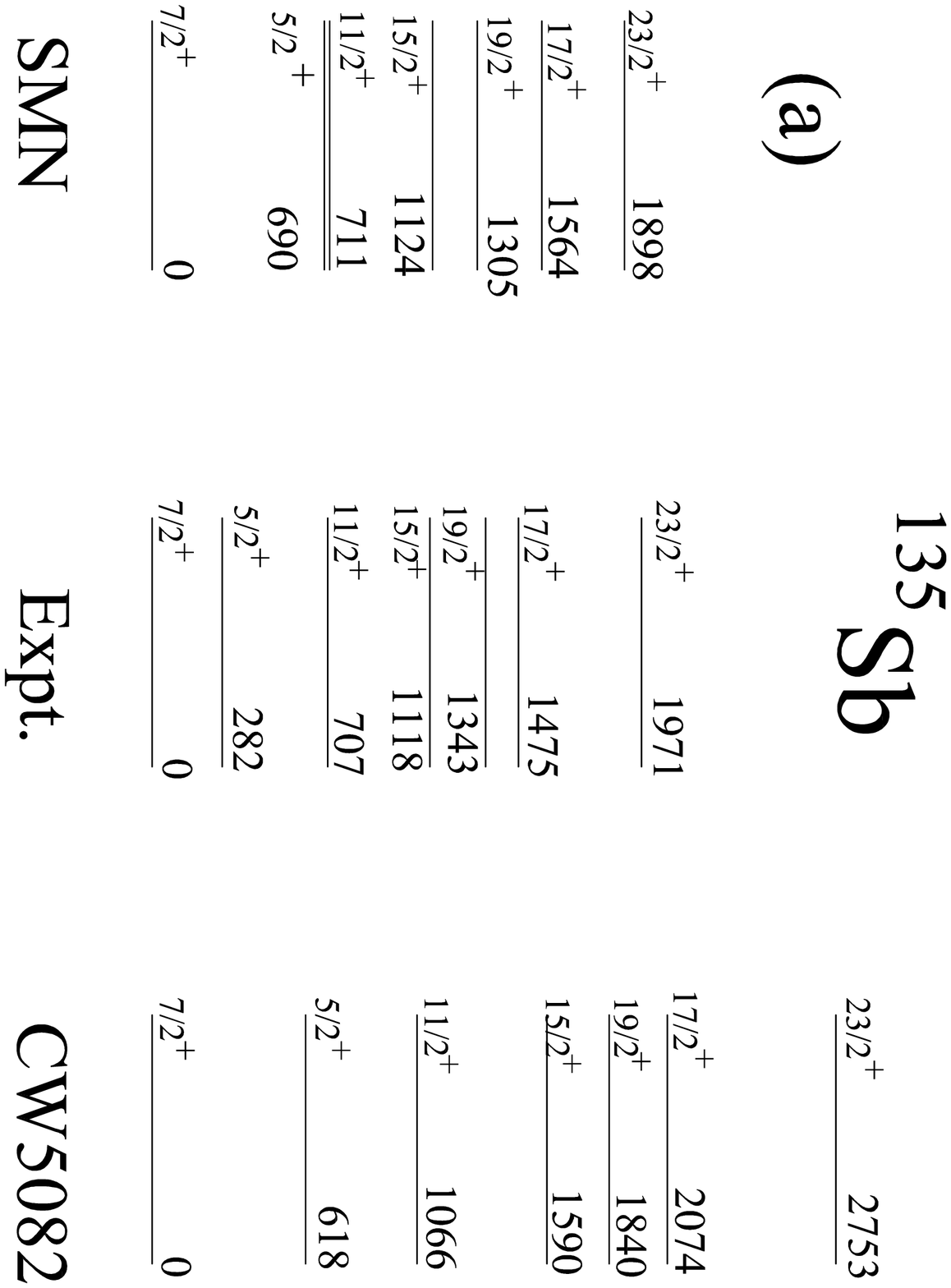}
}
\caption{}
\end{figure*}

\begin{figure*}
\resizebox{0.95\textwidth}{!}{
\includegraphics{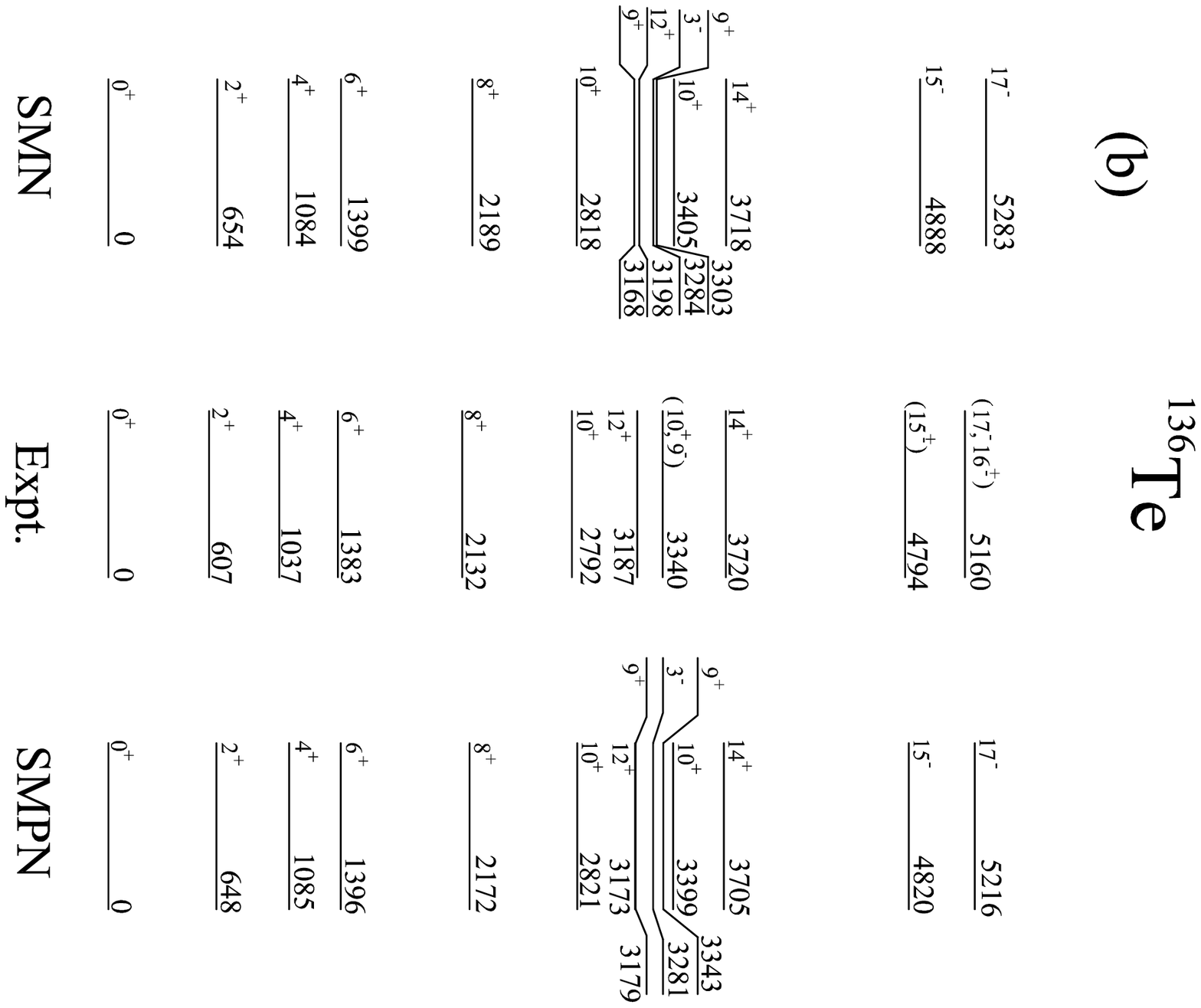}
}
\caption{}
\end{figure*}

\begin{figure*}
\resizebox{0.95\textwidth}{!}{
\includegraphics{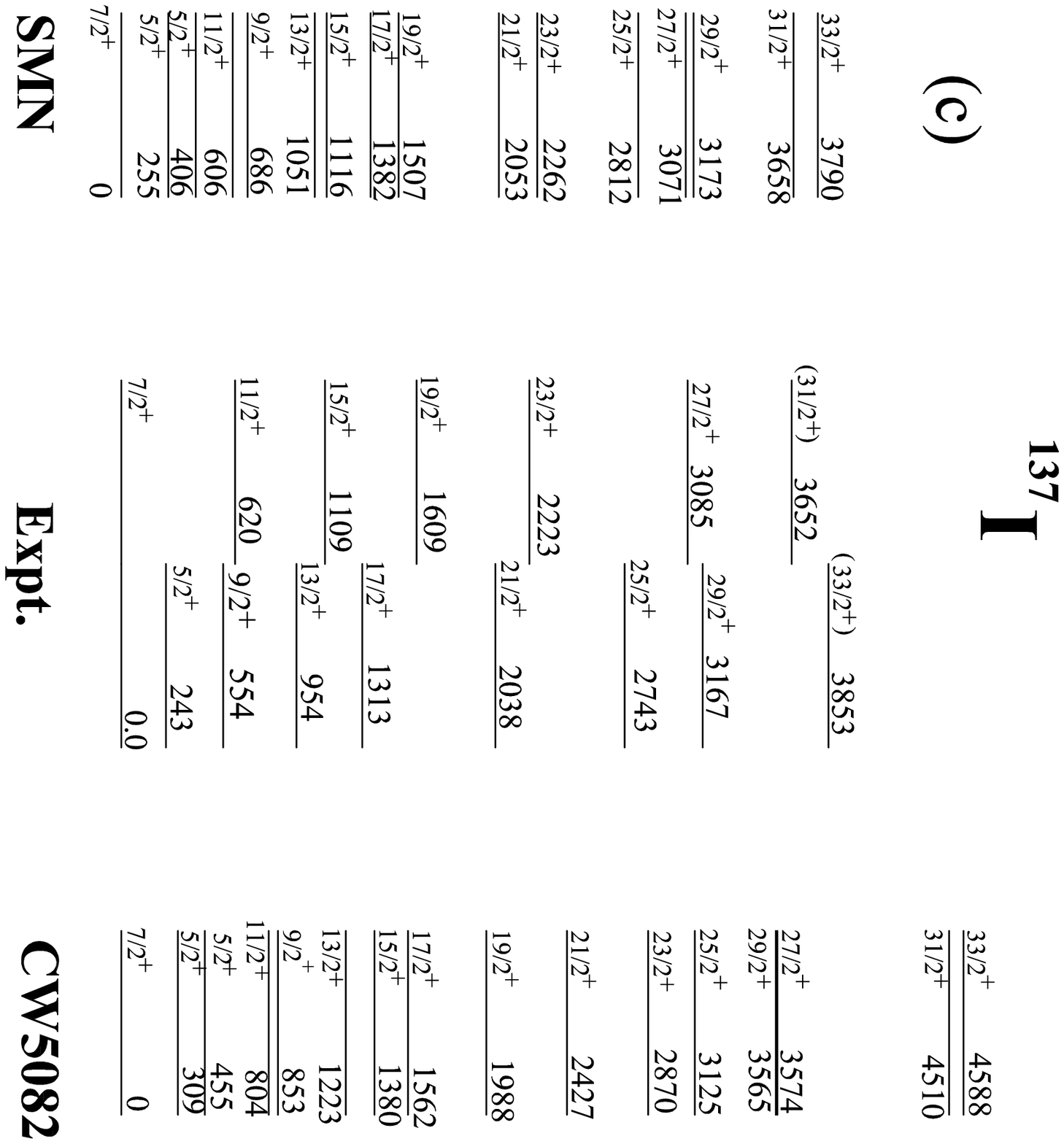}
}
\caption{}
\end{figure*}

\begin{figure*}
\resizebox{0.95\textwidth}{!}{
\includegraphics{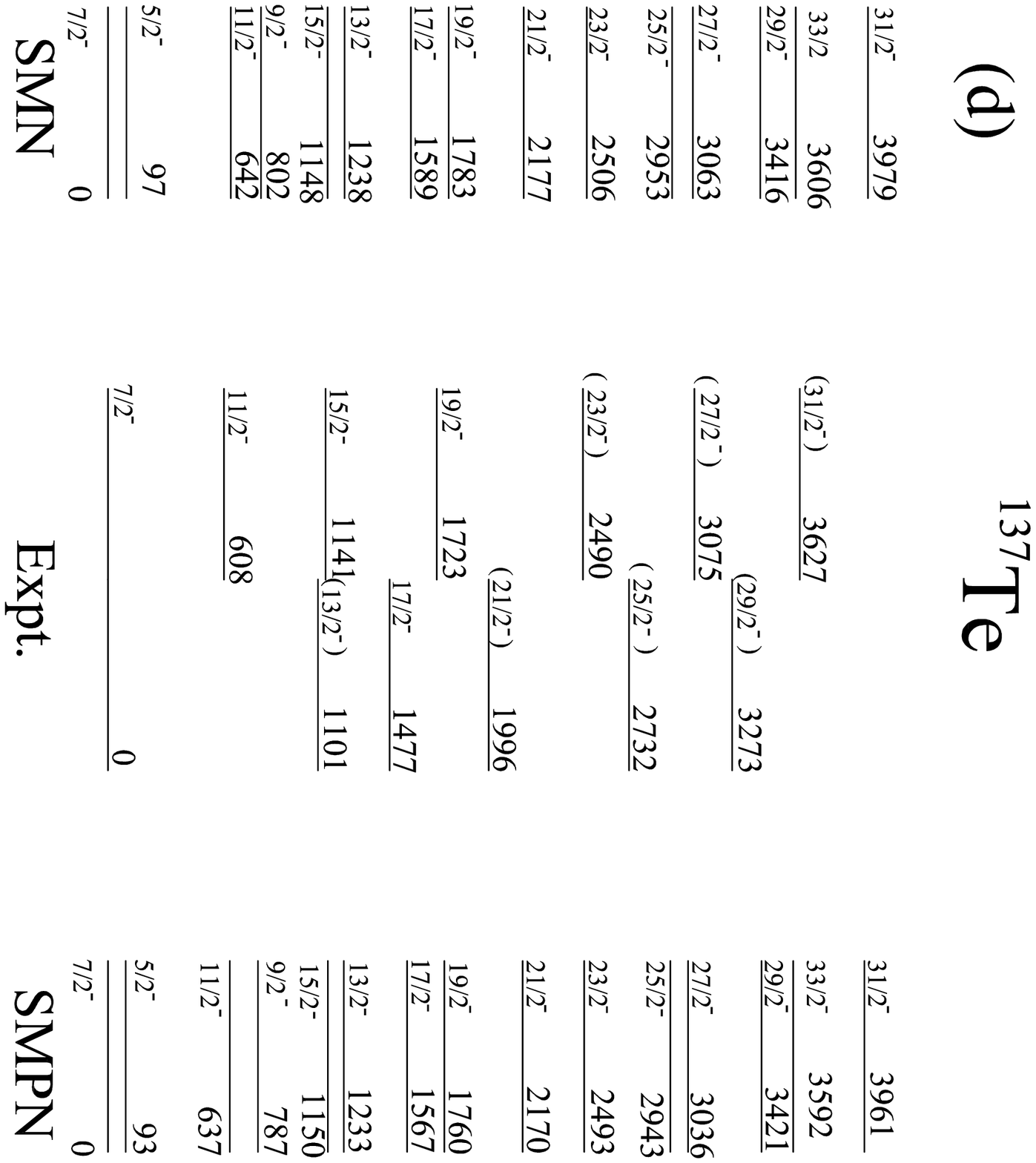}
}
\caption{}
\end{figure*}
 \end{document}